\RequirePackage{fix-cm}

\documentclass[twocolumn]{article}    

\usepackage{authblk}
\usepackage{mathptmx}     
\usepackage{graphicx}
\usepackage{amsmath}
\usepackage{mathtools}
\usepackage[numbers]{natbib}
\usepackage{color}
\definecolor{darkyellow}{RGB}{102,102,0}
\definecolor{orange}{RGB}{255,165,0}
\definecolor{purple}{RGB}{102,0,153}

\begin{document}

\title{Electronic emission of radio-sensitizing gold nanoparticles under X-ray irradiation : experiment and simulations}       

\author[1]{R.Casta \thanks{Electronic address: \texttt{romain.casta@irsamc.ups-tlse.fr}}} 
\author[1]{J.-P.Champeaux}
\author[1]{M.Sence}
\author[1]{P.Moretto-Capelle}
\author[1]{P.Cafarelli}
\author[1]{A.Amsellem}
\author[2]{C.Sicard-Roselli}

\affil[1]{Laboratoire Collisions Agr\'{e}gats R\'{e}activit\'{e}, IRSAMC, CNRS, UMR 5589, Universit\'{e} de Toulouse, UPS, F-31062 Toulouse, France.
              }
              
\affil[2]{Laboratoire de Chimie Physique, CNRS UMR 8000, Universit\'{e} Paris-Sud 11, Bât.350, 91405 Orsay Cedex, France.
}

\date{March 25,2014}
\maketitle

\begin{abstract}
In this paper we present new results on electronic emission of Gold Nanoparticles (GNPs) using X-ray photoelectron spectroscopy (XPS) and compare them to the gold bulk electron emission. This subject has undergone new interest within the perspective of using GNPs as a radiotherapy enhancer. 
The experimental results were simulated using various models (Livermore and PENELOPE) of the Geant 4 simulation toolkit dedicated to the calculation of the transportation of particles through the matter. Our results show that the GNPs coating is a key parameter to correctly construe the experimental GNPs electronic emission after X-ray irradiation and point out some limitations of the PENELOPE model. Using XPS spectra and Geant4 Livermore simulations,we propose a method to determine precisely the coating surface density of the GNPs. We also show that the expected intrinsic nano-scale electronic emission enhancement effect - suspected to contribute to the GNPs radio-sensitizing properties - participates at most for a few percent of the global electronic emission spectra of the GNPs compared to gold bulk.   
\end{abstract}

\section{Introduction}

For many years it has been observed that high-Z materials can cause significant tissue damages when they are coupled with X-Ray radiations \cite{Rosengren1,Rosengren2}. The idea of using these materials properties in cancer therapy has gained interest but important limitations have appeared like cancerous cell targeting and toxicity. Gold Nanoparticles (GNPs) seem to overcome these difficulties because of their supposed non-toxicity \cite{Vujacic} and their capability to enter tumor cells \cite{Herold}. 
Thus several studies on physical and biological GNPs properties have been recently undertaken. Biological studies have shown an important enhancement of the surviving rate on mice treated with X-ray radiations \cite{hainfeld2004,Herold} combined to GNPs, whereas GNPs  without X-Ray have no effect on  tumor cells. This raises the question of the physical properties causing damages to cancerous cells when they interact with an ionizing radiation. These properties can be for example: hyperthermia causing cell death  \cite{Kennedy}, radical production \cite{Carter} or electron emission. 

Concerning this last property, one of the hypothesis \cite{Sanche} is that GNPs can cause damage to DNA via low-energy electron emission. This is supported by experimental studies showing an enhancement of the DNA breaks by GNPs \cite{Butterworth} and the low-energy electron capability to break DNA \cite{Brun_Damage_DNA_LEE,Brun3} but also by theoretical studies showing the electron irradiation dose enhancement by GNPs \cite{Belfast,Jones}.
But, this theory suffers from an important lack of experimental data. Indeed there are very few experimental results available \cite{Xiao} about electron emission of GNPs undergoing X-Ray radiation.
In this paper, we will present and discuss experimental electronic emission spectra of GNPs performed at X-Ray photo-electron spectroscopy (XPS) facility of ENSIACET-CIRIMAT (Toulouse, France), and compare them with the gold bulk spectrum. These experimental results will be compared to Geant 4 simulations of irradiated citrate-coated and uncoated GNPs performed with  Livermore and PENELOPE models \cite{geant41,geant42,penelope,PhysicG4}. 

\section{Experimental Methods}

The GNPs used in our experiment were prepared using the Turkevich method \cite{Turkevitch} which produces GNPs coated with a citrate (${C_6H_5O_7}^{3-}$) monolayer. GNPs were washed by three centrifugation cycle as described by Brun et al. \cite{Brun3}, in order to remove most of the citrate and the  chemical reactants. After washing, the GNPs measured radius is $16nm$. This is a mean radius computed with the software ImageJ from radii measured over four hundred nanoparticles on scanning electron microscope images.

\begin{figure}
\includegraphics[width=0.4\textwidth]{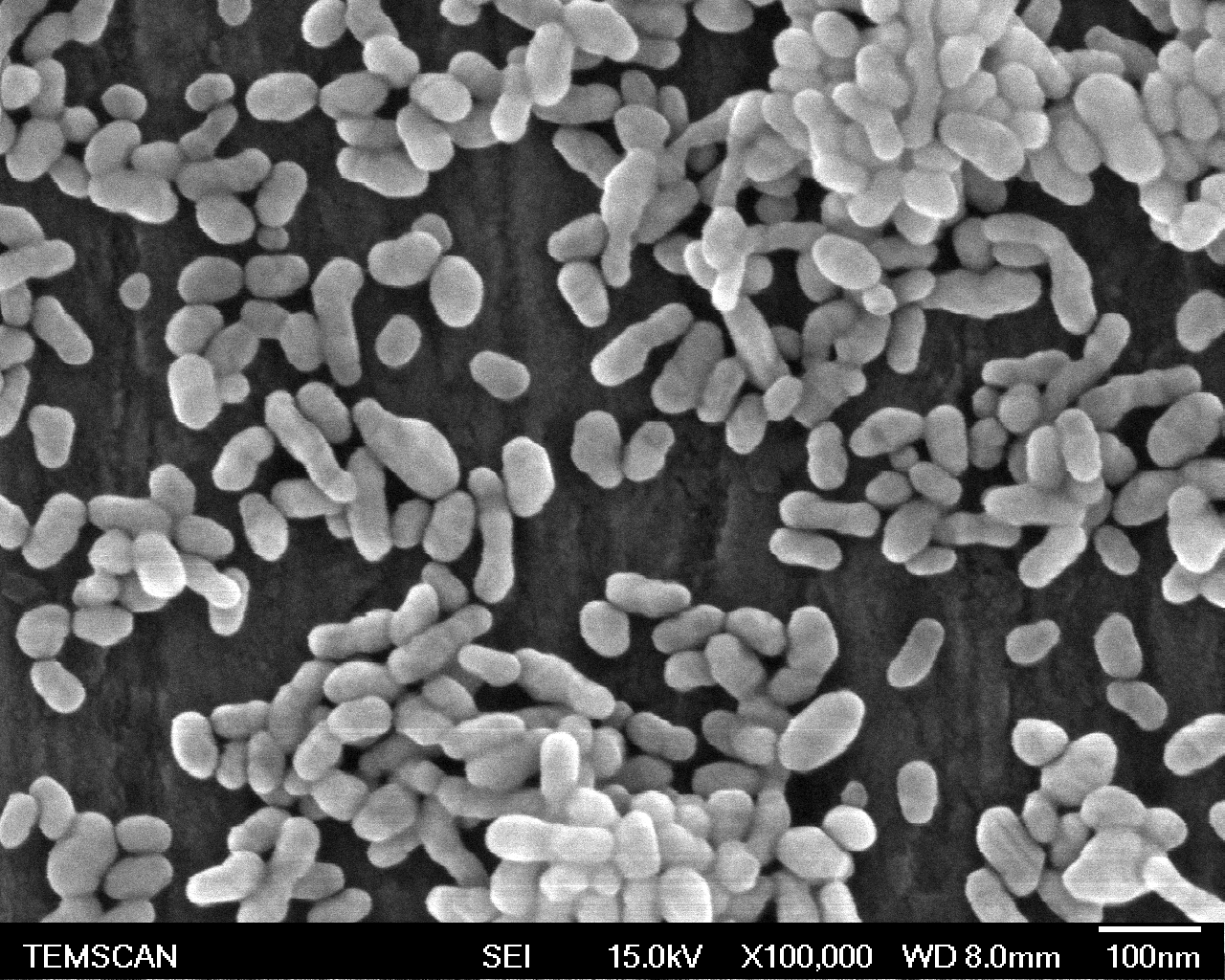}
  \caption{Electronic microscope image of GNPs on aluminum substrate. The size of the  surface shown on the picture is $1200nm \times 900 nm$.}
    \label{fig:Fig1}
 \end{figure}

The initially spherical $16nm$ radius GNPs amalgamate with time from the moment they are produced at the Laboratoire de Chimie Physique (LCP) to their use in our experiment. The GNPs are deposited on an aluminum substrate and characterized using scanning electron microscope. The corresponding image is shown in Fig.~\ref{fig:Fig1}. By analysing the microscope images with the software ImageJ, we were able to determine that GNPs finally have a $19nm$ mean radius. This result will be used in the Geant 4 simulations. 
    
According to Toma at al. \cite{Toma} the GNPs should be inter-spaced with a distance superior to five times their own radius to prevent interactions between themselves. But due to the used deposition technique, the GNPs density is not well controlled and varies from one side to an other on the substrate. In order to reduce possible collective effects between GNPs we have chosen to focus on a low density region of the substrate. 

\begin{figure}
\includegraphics[width=0.4\textwidth]{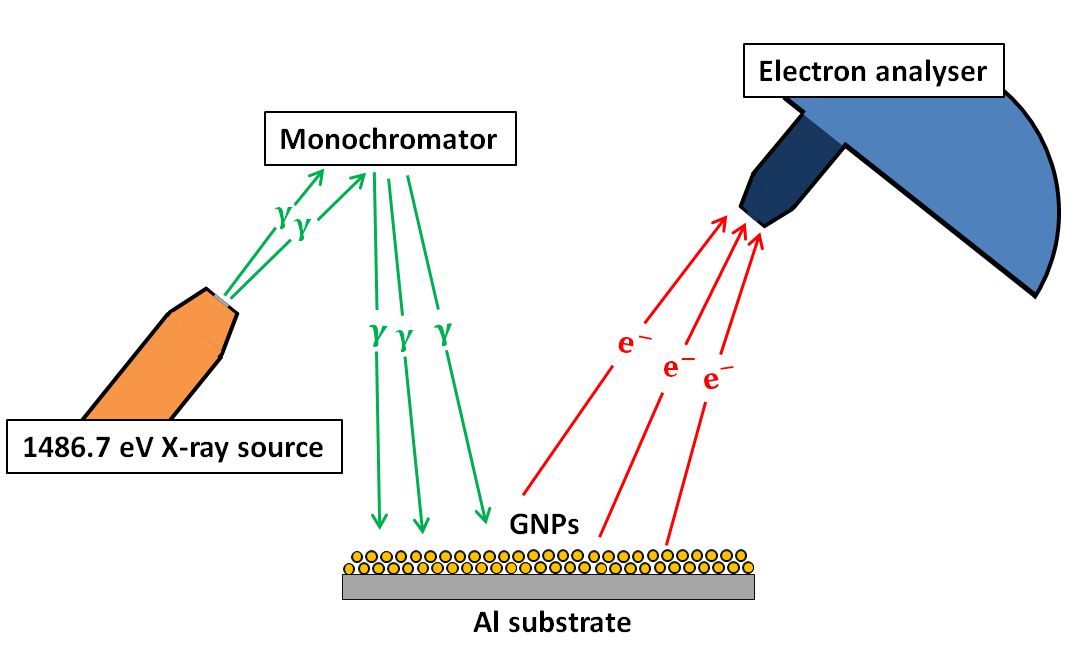}
  \caption{Principle scheme of the experiment performed with the Thermo Scientific K-alpha spectrometer. }
    \label{fig:Fig9}
 \end{figure}

The XPS analysis was performed on gold bulk, on aluminum substrate alone and on GNPs deposited on aluminum substrate with the ENSIACET-CIRIMAT XPS system. This system (Thermo Scientific K-Alpha) represented on Fig.~\ref{fig:Fig9} is a fully integrated spectrometer  using a monochromatic aluminum K-alpha X-Ray source at $1486.7 eV$ coupled with a $180^o$ double focusing hemispherical analyzer having a resolution below $10 meV$ in a wide electron energy range ($100 eV$ to $1.5 keV$). The apparatus is maintained to ultra high vacuum at a pressure lower than $5 \times 10^{-9} mbar$.

From this apparatus we measure the electron spectra in the energy range $136 eV-1487 eV$ with a size of bin of $1 eV$. The low limit energy is imposed by the focus lens of the hemispherical analyzer which works at $100 eV$. 
GNPs spectrum were deduced by subtracting the substrate spectrum from the total (GNPs plus substrate) spectrum. GNPs and gold bulk spectra were normalized to the integral of the first observed spectral line at $1480 eV$ for each spectrum. 
The choice of this line for normalization is motivated by the fact that  it is well defined and its baseline is not perturbed by background induced by scattered electrons as observed for other gold spectral lines at lower kinetic energies.

The X-ray energy available for this experiment was lower than the ones used in medical X-ray sources which can go from few tens of $keV$ to few $MeV$, because the large energy distributions of the X-ray medical sources do not allow the spectra  analysis done in this paper with a highly monochromated source resulting in highly resolved electron spectra which are more interesting to compare to simulations results.
 Nevertheless the physical processes involved in this experiment (photo-electric process, electronic scattering and ionisation) are mainly the ones involved in an experiment at larger X-ray energy. That is why an understanding of this experiment can greatly help to understand the physics involved in medical X-ray  radiotherapy. 

\section{Geant4 Simulations methods}

In order to simulate our XPS measurements on GNPs and gold bulk, we use PENELOPE (PENetration and Energy LOss of Positrons and Electrons) and Livermore models \cite{penelope,PhysicG4} which include among others the photoelectric process, electrons scattering and electron impact ionization process. In our case, these models are implemented in the transport toolkit Geant 4 \cite{geant41,geant42} that we use in its 4.9.6 version. We have  simulated the particle trajectories until $100eV$ both with Livermore and PENELOPE model.
In these simulations, the geometries consist of stacks of few GNPs layers (one to five). These layers are composed of ten thousand $19nm$-GNPs uniformly distributed on a $3 \mu m \times 3\mu m$ surface. A gap of $1 nm$ has been set  between GNPs layers i.e. much lower than a GNP radius in order to take into account the electronic GNPs interactions and to simulate the GNPs piles observed on microscopic pictures (Fig.~\ref{fig:Fig1}). The gold bulk geometry consists of a $3 \mu m \times 3\mu m \times 0.1 \mu m$ cuboid. Despite the fact that both GNPs and the gold bulk cuboid are composed of gold bulk, we will call the gold cuboid "gold bulk" in the further sections.

Each simulation has been done for citrate coated and uncoated GNPs. The citrate coating is simulated by a homogeneous material circumposing the GNPs, composed of $C$, $H$ and $O$ in citrate stoichiometric proportions ${C_6H_5O_7}^{3-}$ and with a density of $1 g.cm^{-3}$. The density for this material has been chosen considering the fact that the GNPs external layer is not only composed of citrate ($1.7 g.cm^{-3}$ for sodium-citrate) but also of water and ion shells. Simulation have been done using $1030$ molecules/GNP in a $71pm$ thick coating shell which has been chosen in order to fit the experimental results. This fitting process will be explained further in this paper. 

\begin{figure}
   \includegraphics[width=0.4\textwidth]{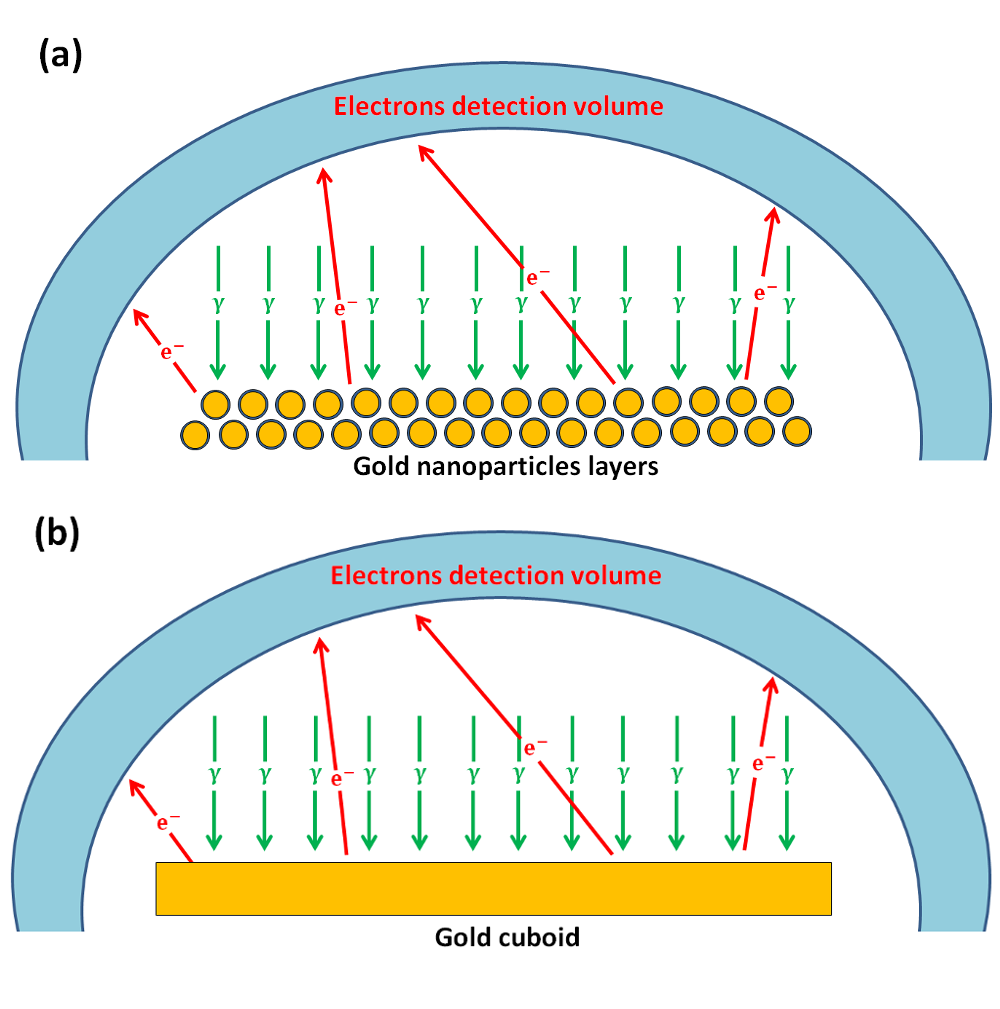}
  \caption{Geant4 simulation geometries for GNPs layers \textbf{(a)} and gold cuboid \textbf{(b)}.\label{fig:Fig10}}
 \end{figure}

GNPs layers (coated and uncoated) and gold bulk surfaces are perpendicularly irradiated by four hundred million $1486.7 eV$ photons, uniformly distributed in position on a $309 nm$ side square centered on GNPs layers and gold bulk to avoid undesirable edges effects. Indeed, in regard to the size of the incident photons spot compared to the size of both GNPs layers and gold bulk, we can consider these last ones as infinite surfaces. The simulation geometry is schematically represented on Fig.~\ref{fig:Fig10}. Energies of all electrons emitted out from the upper part of GNPs stacks or gold bulk are recorded. From the four hundred million photons occurrences we have collected between four and seven million electrons, a number that is statistically relevant to compare electron energy spectra to the experimental ones.

\section{Experimental results}

 \begin{figure}
   \includegraphics[width=0.4\textwidth]{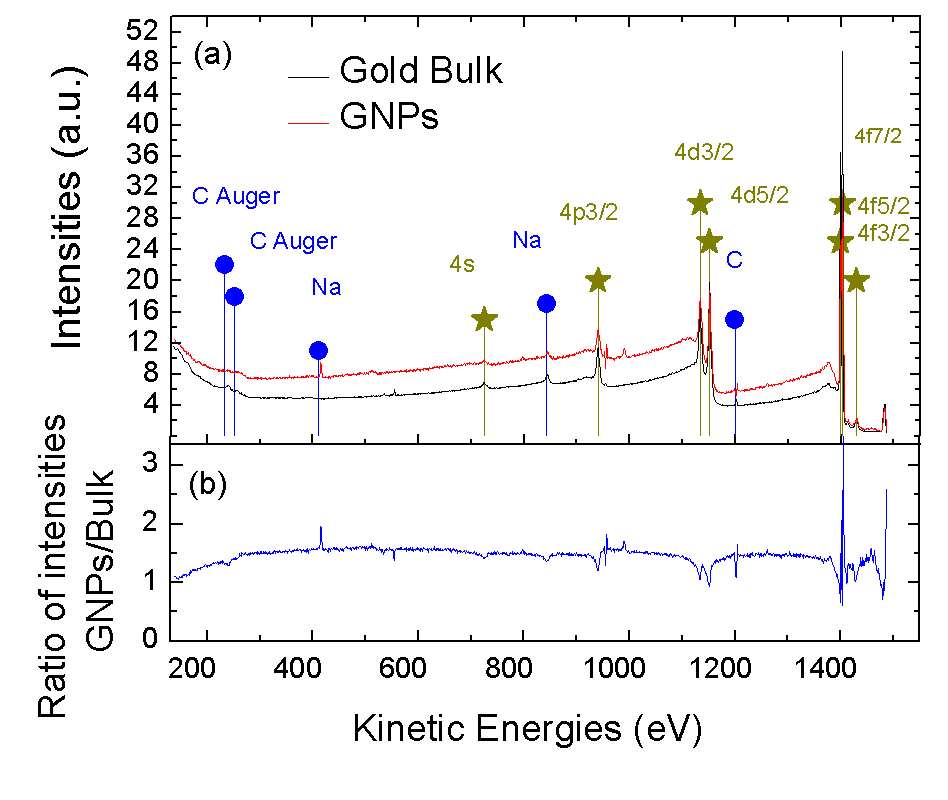}
  \caption{Electrons energy spectra obtained by $1486.7 eV$-XPS analysis for both gold bulk and citrate-coated GNPs on the range $136-1550 eV$ with $1eV$ bin (a) and the ratio GNPs/Bulk of these spectra (b). Vertical lines point out the spectral gold lines ($\textcolor{darkyellow}{\star}$) and the contaminants lines ($\textcolor{blue}{\bullet}$).\label{fig:Fig2}}
 \end{figure}

The experimental XPS spectra and ratio of citrate-coated GNPs and gold bulk are presented on Fig.\ref{fig:Fig2}. The atomic gold spectral lines are well identified. We observe that the atomic gold photo-electron peaks are superimposed on an electronic scattering background. 
Each photo-electron peak exhibits the same shape, a small scattering peak following the main gold photo-electric peak a few $eVs$ after. Indeed, the electrons of the main atomic lines come from the samples surfaces (gold bulk or GNPs) and are not scattered. They go to the detector without any energy loss, whereas the gold photo-electrons extracted from deeper part of the GNPs stacks or gold bulk, are scattered and give rise to a secondary peak and a scattering tail a few $eV$ before the main peak. 

\begin{table*}
\begin{tabular}{ccccc}
\hline\noalign{\smallskip}
  Line &NIST&Experiment&Spectral line integral&Ratio \\
  &&$\pm FWHM$&GNPs - bulk&GNPs/bulk \\
  &$eV$&$eV$&$(a.u.)\times eV$\\
  \hline\noalign{\smallskip}
  $5p_{3/2}$ & 1429.4 & $1428\pm6$ &  5.67 - 4.20   & 1.35\\
  $4f_{7/2}$ & 1402.6 & $1401\pm4$ & 85.05 - 78.45  & 1.08\\
  $4f_{5/2}$ & 1398.8 & $1398\pm2$ & 52.73 - 66.74  & 0.79\\
  $4f$       & -      & -          & 137.78 - 145.2 & 0.95\\
  $4d_{5/2}$ & 1151.4 & $1150\pm5$ & 63.40 - 79.56  & 0.80\\
  $4d_{3/2}$ & 1133.4 & $1132\pm6$ & 51.50 - 55.87  & 0.92\\
  $4d$       & -      & -          & 114.9 - 135.43 & 0.85\\
  $4p_{3/2}$ & 940.1  & $939\pm6$  & 19.86 - 25.12  & 0.79\\
  \hline\noalign{\smallskip}
\end{tabular}
\caption{\label{Tab:Tab1}Experimental X-ray line table for citrate-coated GNPs and gold bulk. For each spectral line, it shows the energies from the NIST database \cite{NIST}, our experimental values, the integral spectral line value and the intensities ratio between GNPs and gold bulk. The $4f$ and $4d$ lines are the sum of the integral of the peaks $4f_{7/2}$,$4f_{5/2}$ and $4d_{5/2}$,$4d_{3/2}$.}
\end{table*}

Tab.~\ref{Tab:Tab1} presents for both GNPs and gold bulk the energy position of each experimental photo-electrons peak compared to the NIST \cite{NIST} XPS database values. We calculated the integral for each experimental peak after subtracted its baseline. The ratios of these integrals for GNPs and gold bulk are given in Tab.~\ref{Tab:Tab1}.

For a given spectral line the ratio of the scattering to the main peak (baseline subtracted) is always more important for GNPs than for gold bulk, as observed for the lines $4f$ (GNPs: $0.118$, Bulk: $0.018$), $4d_{3/2}$ (GNPs: $0.067$, Bulk: no scattering peak), $4p_{3/2}$ (GNPs: $0.105$, Bulk: no scattering peak).

The main other difference between GNPs and gold bulk spectra comes from the electronic scattering background which is always higher for GNPs than for gold bulk. Indeed, electrons from the continuum of the spectra can come from scattered photo-electrons inside the material but also from secondary electrons induced by the initial photo-electrons, the amount of these strongly depending on the geometry. The ratio of the intensities between GNPs and bulk is presented on Fig.~\ref{fig:Fig2}(b). This ratio is roughly constant around $1.47$ on the scattering tail and in the range $300-1500 eV$.

In Tab.~\ref{Tab:Tab1}, we observe that the measured ratios are as expected slightly below $1$: the bulk lines intensities are always higher than the GNPs ones and are much lower than the $1.47$ average ratio observed for the electronic background. This can be explained by the fact that the low-energies electrons of these low-energies spectral lines are more absorbed by the citrate-shell than the highest spectral line used for the normalization. The energy absorbed by the citrate shell is redistributed in the electronic emission background composed of secondary and scattered electrons. Thus, there is a decrease of the number of electrons in the spectral lines and an increase of the electron number of the electronic background.

In the $136-300 eV$ range, we observe that the ratio goes slightly up from $1$ to $1.47$ due to geometries of gold bulk and GNPs. This will be developed in a further paper.

\section{Geant4 Simulation results}

\begin{figure}
    \includegraphics[width=0.4\textwidth]{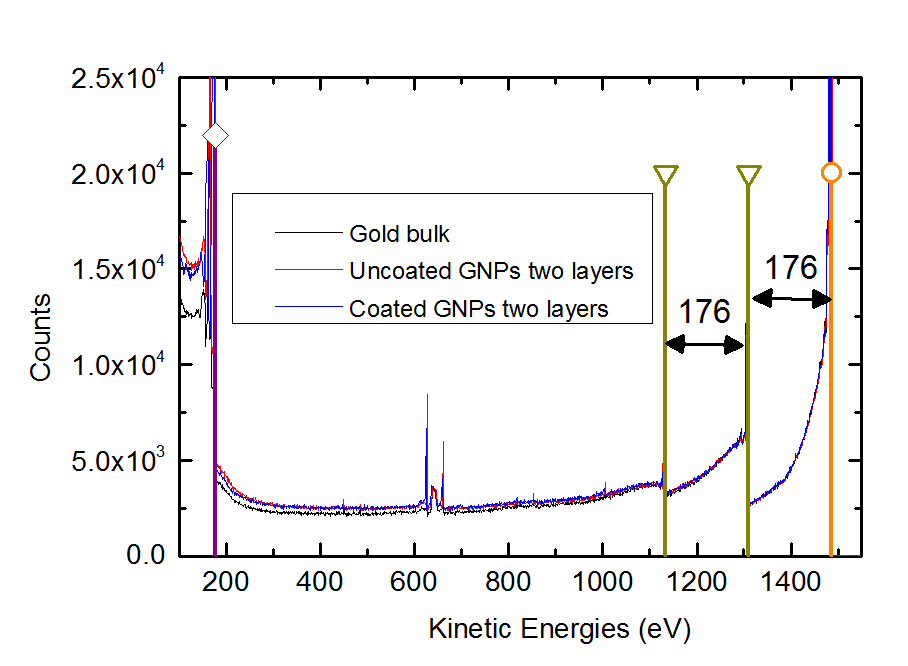}
    \caption{PENELOPE-Geant4 simulated energy spectra ($100-1550eV$) for two citrate-coated GNPs layers stack (blue), two uncoated GNPs layers stack without coating (red) and gold bulk (black). Labels: Photoelectric effect ($\textcolor{red}{\bigcirc}$), energy losses ($\textcolor{darkyellow}{\bigtriangledown}$) and electron ionization impacts ($\textcolor{purple}{\diamond}$). }
    \label{fig:Fig3}
\end{figure}

We present the energy spectra of two coated and uncoated GNPs layers stacks and gold bulk simulated with PENELOPE-Geant4 on Fig.~\ref{fig:Fig3} and with Livermore-Geant4 on Fig.~\ref{fig:Fig4}. We do not show the whole spectral range ($100eV-1500eV$) in this last figure because all the peaks are in the $700eV-1500eV$ range.

\subsection{Discussion on PENELOPE-Geant4 Spectral lines}

In the PENELOPE-Geant4 simulations of the photoelectric process, the atomic external shells above the L shell are not energetically considered. To account for this shell we assign to each photo-electron emitted from shells of atomic number $n \geq 2$, the whole incident photon energy i.e $1486.7 eV$. As a consequence in the $100-1500 eV$ kinetic energy range, PENELOPE is not able to simulate the photo-electrons emission lines observed experimentally for gold bulk and GNPs. And the electrons produced by direct photoelectric process (from atoms of the surface without scattering) only contribute to the first main peak at $1486 eV$ (orange dot) in our simulation. 

Even if the photoelectric process is practically ignored at these low energies, the inelastic scattering and electron impact ionization processes are still computed. 

The simulated peak at $176 eV$ (purple diamonds) corresponds to secondary electrons induced by electron impact ionization and emitted by atoms on the surface but is not equal to a ionization energy of a  gold shell. In PENELOPE model ionization of a given atomic shell is approximated as a single resonance (a $\delta$ distribution) at an energy $W_k$ function of a gold shell ionization energy. We can get the expression of $W_k$ from \cite{penelope}:
\begin{equation}
W_k=\sqrt{(aU_k)^2+\frac{2}{3} \frac{f_k}{Z} \Omega_p^2}
\end{equation}
where $a$ is an empirical adjustment factor, $U_k$ the ionization energy, $f_k$ the number of electrons in the k-th shell and $\Omega_p$ the plasma energy corresponding to the total electron density in the material. Further explanation are available in Geant4 Physic Manual \cite{penelope} in Chapter three "Electrons and positrons interactions", section "Inelastic collisions". 

The signature of these electron impact ionizations is observed at $1310 eV$ and $1134 eV$ (yellow triangles). The $1310 eV$ peak corresponds exactly to a $176 eV$ energy loss from an initial $1486 eV$ photo-electron and the second peak at $1134 eV$ to two successive energy losses caused by two electron impact ionization processes from initial $1486 eV$ electrons. 

\subsection{Discussion on Livermore-Geant4 Spectral lines}

\begin{figure}
    \includegraphics[width=0.4\textwidth]{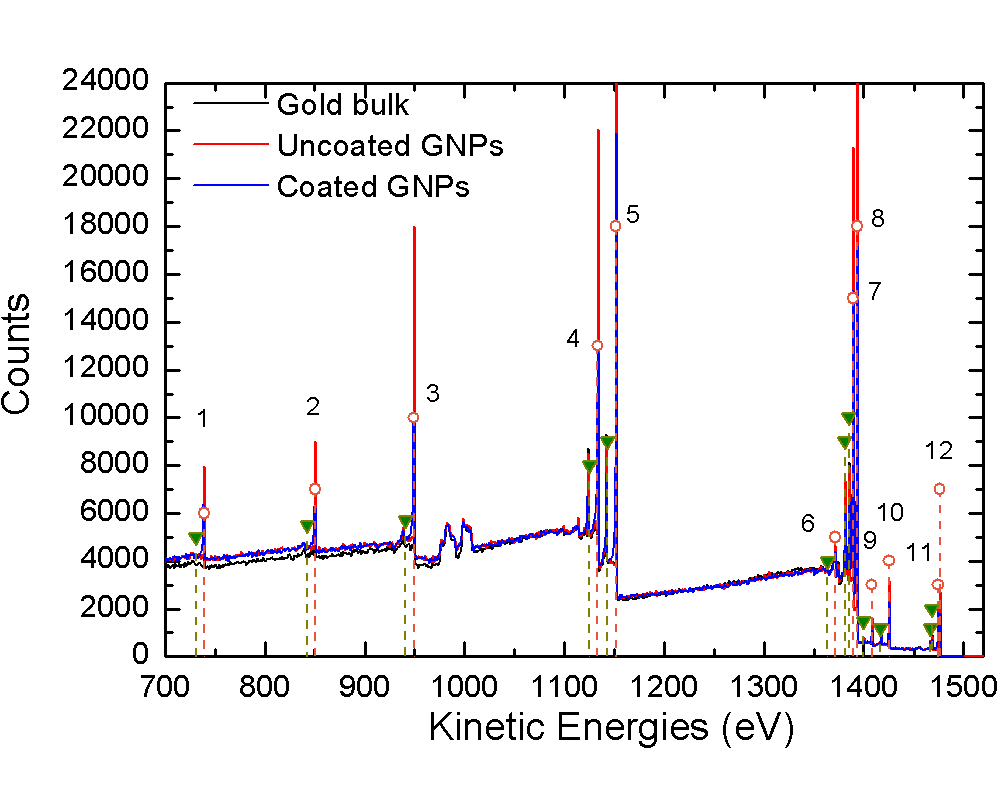}
    \caption{Livermore-Geant4 simulated energy spectra ($700-1550eV$) for two citrate-coated GNPs layers stack (blue), two GNPs layers stack without coating (red) and gold bulk (black). Labels: Photoelectric process peaks ($\textcolor{red}{\bigcirc}$) and energy losses peaks ($\textcolor{darkyellow}{\bigtriangledown}$). }
    \label{fig:Fig4}
\end{figure}

The spectra obtained by using the Livermore Geant4 simulation are shown Fig.~\ref{fig:Fig4}. We can observe two kinds of spectral lines. The photo-electric spectral lines (orange dots) are  much more numerous than in the previous spectra simulated with PENELOPE. Their corresponding energy loss spectral lines are shown (green triangles).
The presence of more numerous photo-electric spectral lines can be explained by the fact that in this model the photo-electric process is handled in a very different way than in the PENELOPE model. The energy levels come from EADL database \cite{EADL} but are taken into account even for the external shells and the photo-electric cross-sections are not computed but interpolated from the tabulated database EPDL97 \cite{EPDL97}. This photo-electric process implementation allows the simulation of all the photo-electric spectral lines predicted by the EADL energy levels except for the one corresponding to the most external shell ($8.3eV$ binding energy). All the photo-electric spectral lines, their intensities and locations compared to NIST values are summarized in Tab.~\ref{Tab:Tab2}. In this table we can observe that the photo-electric peaks are less intense for the coated GNPs spectrum which suggests an absorption of the primary photo-electron by the citrate coating.  
The energy loss spectral lines correspond to the electron-impact ionizations of the most external shell ($E_b=8.3eV$) by the photo-electrons  thus explaining their location at $8.3eV$ before each of the most intense photo-electric spectral lines (apart for the $1486.7eV$ one). The electron-impact ionization process is handled in a similar way than the photo-electric process and the cross-sections are interpolated from the tabulated values of the EEDL \cite{EEDL} database.

By comparing both simulations, we see that the Livermore model is more suited to simulate our experiment than the PENELOPE model. Consequently, this model has been used in our study.

\begin{table*}
\begin{tabular}{ccccc}
\hline\noalign{\smallskip}
  N$^o$  &  NIST   & Livermore  &  Integral                  & Integral ratio \\
         &  Values & simulation $\pm FWHM$ &  coated - uncoated - bulk  &  Coated GNPs/Bulk \\ 
         &         &            &                            &- Uncoated GNPs/Bulk \\ 
         & $eV$ & $eV$ & $counts \times eV$ & \\ 
\hline\noalign{\smallskip}

 1 & - & $739.5 \pm 0.50$ & 2275.29 - 3861.46 - 3805.22 & 0.60 - 1.01\\
 2 & - & $850.5 \pm 0.50$ & 1874.77 - 4559.73 - 4783.48 & 0.39 - 0.95\\
 3 & 940.1 & $949.5 \pm 0.50$ & 5858.84 - 13409.74 - 13242.79 & 0.44 - 1.01\\
 4 & 1133.4 & $1133.5  \pm 0.50$ & 8833.83 - 17316.43 - 16504.98 & 0.54 - 1.05\\
 5 & 1151.4 & $1151.5 \pm 0.50$ &18891.92 - 26709.72 - 25624.89 & 0.74 - 1.04\\
 6 & - & $1371.5 \pm 1.00$ & 1079.50 - 1136.80 - 953.78 & 1.13 - 1.19\\
 7 & - & $1389.5  \pm 0.50$ & 12449.02 - 18705.78 - 17602.84 & 0.71 - 1.06\\
 8 & 1398.8 & $1393.5 \pm 0.50$ & 15874.32 - 23750.61 - 22318.70 & 0.71 - 1.06\\
 9 & 1402.6 & $1408.5 \pm 0.49$ & 621.90 - 1132.88 - 1036.48 & 0.60 - 1.09\\
 10 & 1429.4 & $1425.5 \pm 0.50$ & 1863.79 - 2769.35 - 2596.77 & 0.72 - 1.07\\
 11 & - & $1474.5 \pm 0.50$ & 1132.00 - 1954.00 - 1850.00 & 0.61 - 1.06\\
 12 & - & $1476.5 \pm 0.50$ & 1819.00 - 2646.13 - 2475.94 & 0.73 - 1.07\\
 \hline\noalign{\smallskip}
\end{tabular}
\caption{\label{Tab:Tab2}Livermore simulated rays table for two layers stack coated GNPs, uncoated GNPs two layers stack and gold bulk. For each spectral line are given the energies from the NIST database \cite{NIST} when there is a match, the Livermore simulated values, the integral spectral lines values and the intensities ratios between GNPs and gold bulk.}
\end{table*}

\subsection{Discussion on the electronic background}

\begin{figure}
    \includegraphics[width=0.4\textwidth]{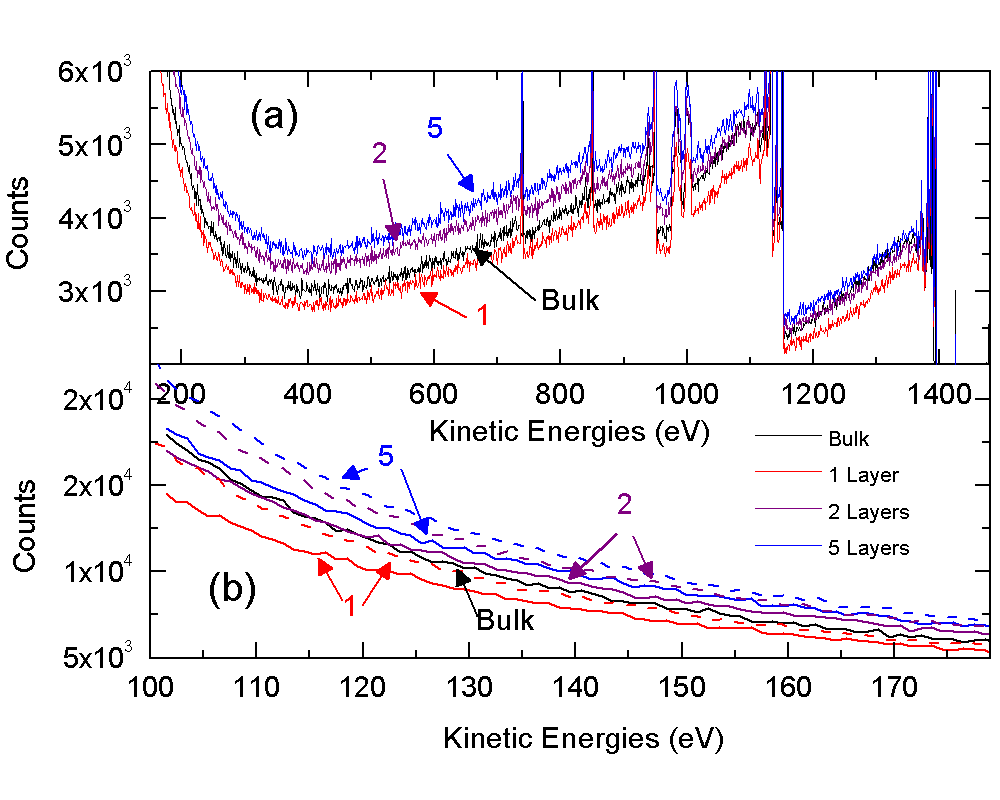}
    \caption{Livermore-Geant4 simulated energy spectra on the range $180-1300eV$ (a) for one (red), two (purple) and five (blue) GNPs layers and gold bulk (black) and on the range $100-180eV$ (b) for one (red), two (purple) and five (blue) citrate-coated GNPs layers stacks, for one (dashed red), two (dashed purple) and five (dashed blue) uncoated GNPs layers stacks and gold bulk (black). }
    \label{fig:Fig5}
\end{figure}

Fig.~\ref{fig:Fig5} shows the emission electronic background of the Livermore-Geant 4 simulations for one, two and five citrate-coated and uncoated GNPs layers stacks as compared to the gold bulk electronic emission. Fig.~\ref{fig:Fig5}(a) shows the spectrum in the range ($180-1500 eV$). In this range of the spectrum  the citrate-coated and uncoated spectra are not distinguishable so we have chosen to present only the citrate-coated results. Fig.~\ref{fig:Fig5}(b) shows the spectrum in the $100-180 eV$ energy range. 

We can observe that the intensities of the electronic background increase from one to five layers. The two and five layers stacks are very close and show clearly a gap with the one layer intensity. The bulk intensity is located between one layer and two layers intensities, close to the one layer curve at low kinetic energy and to the two and five citrate-coated GNPs layers stacks intensities at high energy.

By comparing emission spectra of gold bulk and GNPs layers stacks, and following the idea that a gold bulk can be approximated by an infinity of GNPs layers, one would expect to find the same spectra for bulk GNPs layers stacks after a certain high number of layers, which is not the case at low energies in our simulations. Indeed, the intensities of the electronic background of stacks composed of more than two GNPs layers merged with the gold bulk intensity at high energies ($>900 eV$), but for low energies difference between these emissions spectra are significant.

The relatively important difference observed between one GNPs layer and two GNPs layers stack intensities shows the role played by the multiple electrons scattering between GNPs from the different layers. Indeed some electrons emitted from the deeper layer can impact GNPs of the surface layer with sufficient energy to ionize atoms and produce a secondary electron cascade which will increase the global intensity of the two layers stack relatively to one layer. 
Because these secondary electrons scatter through several GNPs, they have a longer path up to the surface and a lower energy than the electrons from the surface layer. 

\begin{figure}
    \includegraphics[width=0.4\textwidth]{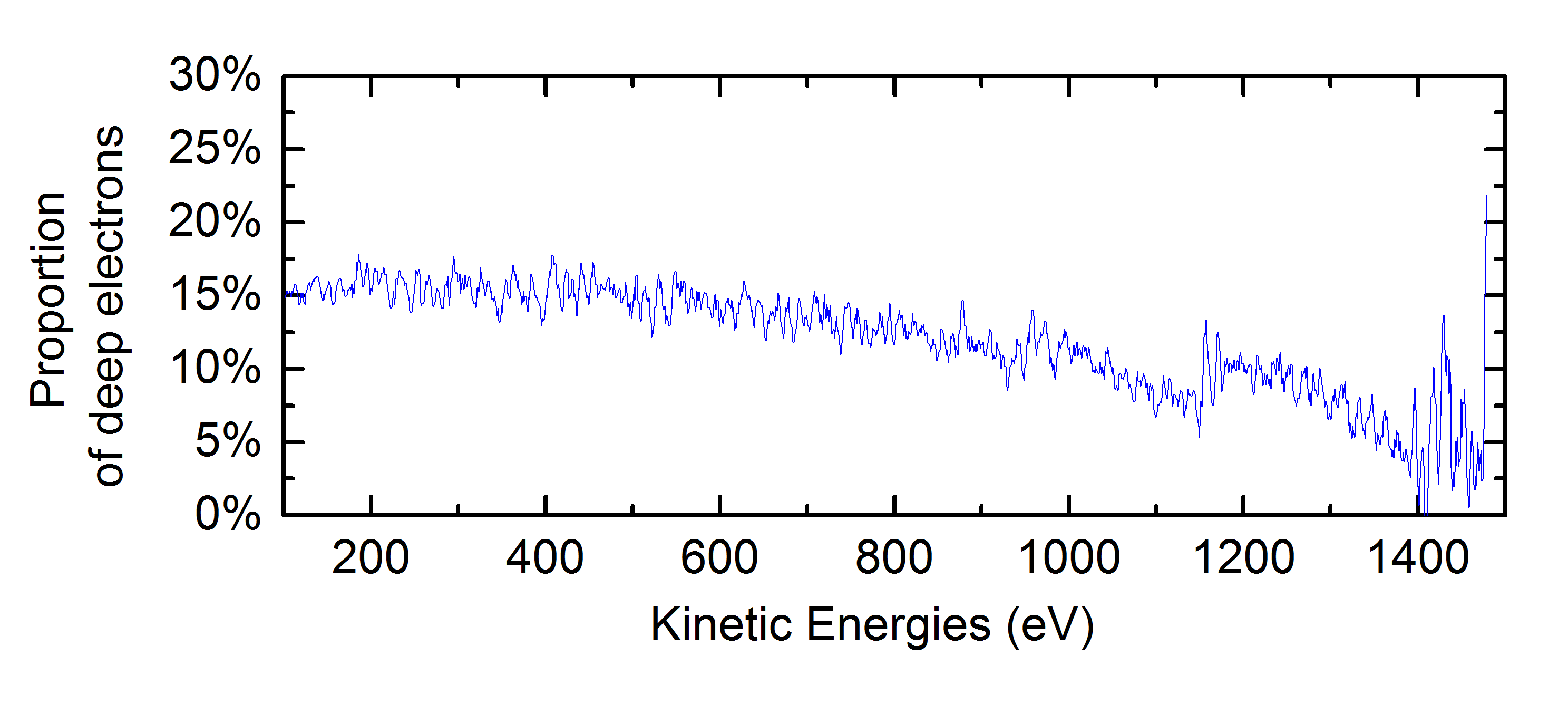}
    \caption{Difference of the signal between two and one layer  stack intensity divided by the two layers stack intensity as a function of the electron energy.}
    \label{fig:Fig6}
\end{figure}

Fig.~\ref{fig:Fig6} represents the difference of the signal between two and one layer  stack intensity divided by the two layers stack intensity as a function of the electron energy i.e. the proportion of electrons generated by the deep layer electronic emission inside the two layers stack intensity. We clearly observe that the lower the energy, the larger the gap between one layer and two layers intensities, varying from $4\%$ at $1486 eV$ up to $15\%$ at $150eV$. This observation confirms our previous explanation of the difference between one and two GNPs layers.

\begin{figure}
    \includegraphics[width=0.4\textwidth]{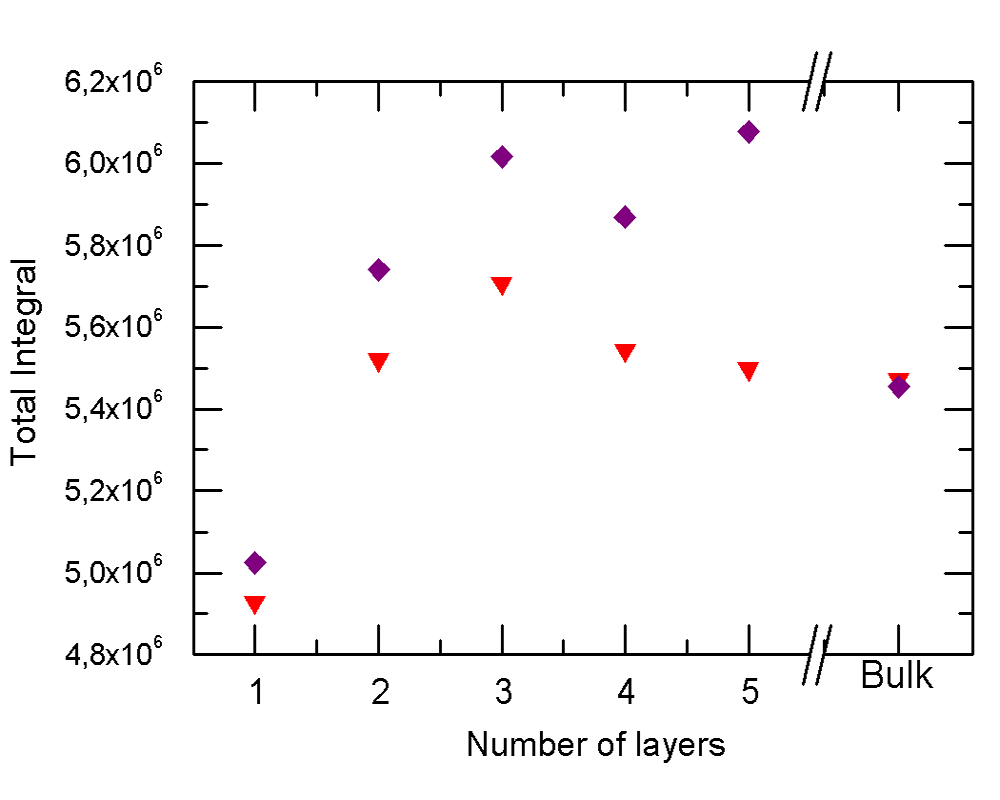}
    \caption{Total integral of kinetic energy spectra simulated by Livermore model for the bulk, one, two, three, four and five citrate-coated ($\textcolor{purple}{\diamond}$) and uncoated ($\textcolor{red}{\bigtriangledown}$) GNPs layers stacks.}
    \label{fig:Fig7}
\end{figure}

Fig.~\ref{fig:Fig7} presents the total spectrum integral as a function of the number of layers: there is an obvious decrease of the variation of the total electronic emission after two layers.

The electrons coming from layers deeper than the second have an always longer path up to the surface and therefore a probability always lower to reach it, thus explaining this decreasing after two layers. 

\section{Comparison Simulation / Experiment}

In order to compare our experimental results to simulated ones, we have chosen the case of two simulated GNPs layers. The choice is driven by the scanning electron microscope picture presented on Fig.~\ref{fig:Fig1} that shows that most of the particles stacks are not higher than two or three GNPs. Moreover, we have seen that the difference between electronic emission layers stack is small for more than two layers.  

\begin{figure}
    \includegraphics[width=0.4\textwidth]{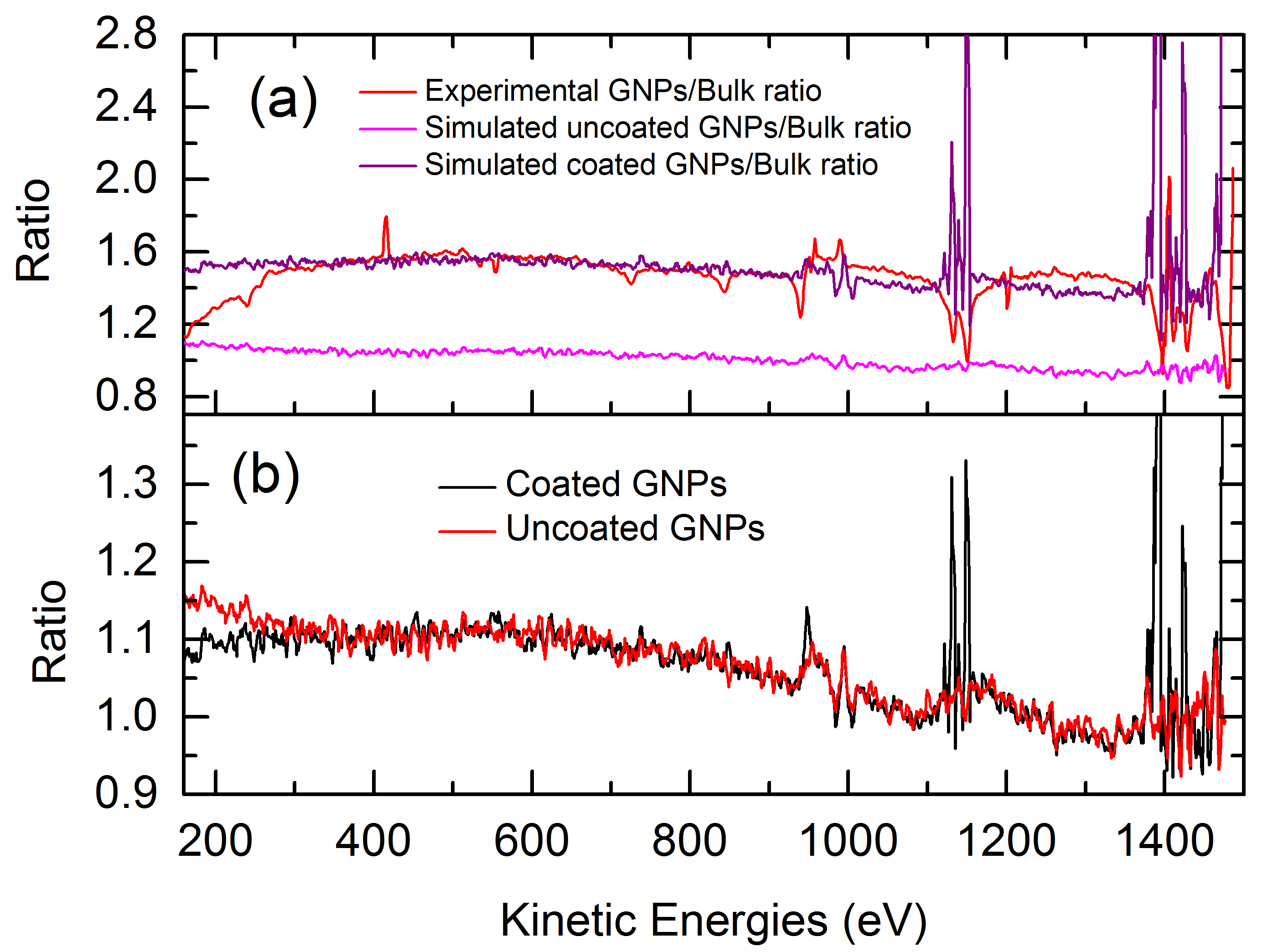}
    \caption{\textbf{(a)} Experimental ratio of two layers stack of GNPs to gold bulk normalized intensity as in Fig.~\ref{fig:Fig2} (red) compared to the Livermore-Geant4 ratios of citrate-coated (purple) and uncoated (pink) GNPs two layers stacks to gold bulk. The simulated spectra were normalized on their energetically highest spectral lines.
    \textbf{(b)}The coated (black) and uncoated (red) two layers stacks GNPs to bulk ratios without normalization.}
    \label{fig:Fig8}
\end{figure}

For this comparison, we chose to focus on the ratio between two GNPs layers stacks and gold bulk spectra. In order to be as close as possible of the experimental spectra, we normalized the two coated and uncoated GNPs layers stacks and the gold bulk spectra on their energetically highest spectral lines. The GNPs/Bulk normalized intensities ratios are presented on Fig.\ref{fig:Fig8}(a) for coated and uncoated GNPs compared with experimental GNPs/bulk ratio. The GNPs/Bulk ratios without normalization are also presented on Fig.~\ref{fig:Fig8}(b).

We observe that above $300eV$ the experimental and  GNPs citrate-coated ratios match around $1.5 \pm 0.1$. The gap between the two ratios increases below $300eV$. The experimental ratio is best fitted by $1030$ citrate molecules per GNP.

As seen from the same ratios without coating this gap is due to the citrate monolayer on GNPs. The comparison of the intensity ratio with and without normalization on Fig.~\ref{fig:Fig8}(a) and  Fig.~\ref{fig:Fig8}(b) shows that this gap is actually due mostly to the normalization method on the highest spectral lines : The ratio without normalization is around $1.10\pm 0.05$ in the range $200-900 eV$ which is much lower than the ratio observed with the normalization.

To explain this difference between the ratios with and without normalization, we need to focus on the highest spectral line intensities of bulk and coated GNPs (last line in Tab.~\ref{Tab:Tab2}). We observe that the citrate coated GNPs spectral line is less intense than the gold bulk one by a factor $0.73$, because the citrate coated GNPs spectral line is reduced by the citrate shell. 
Therefore, to normalize the coated GNPs spectrum on the highest spectral line we have to multiply the intensity of this spectrum by a factor $1/0.73$, and consequently the ratio is around $1.10$ without normalization but takes a value around $1.10/0.73 \simeq 1.50$ with this highest spectral line normalization method. 
Therefore, we can suppose that the actual experimental electronic emission enhancement is only around $1.10$ as the simulated one (Fig.~\ref{fig:Fig8}(b)). As a consequence, if there is an enhancement above $300eV$, it is relatively low - around a few percents - and it is purely due to GNPs, independently of the coating since we observe the same ratios for coated and uncoated GNPs. Below $300eV$ the simulated results do not fit very well the experimental ones and a large gap appears between the two ratios.

We can use these well understood ratios above $300eV$ to deduce the surface citrate molecules density on GNPs. The ratio of the normalized intensities of coated GNPs and gold bulk is very sensitive to the number of citrate molecules per GNP. The same Geant 4 simulations performed with a variation of $50$ citrate molecules (thus modifying the citrate shell thickness) leads to a variation of the ratio of $0.1$. Such a large variation allows us to adjust the citrate molecules number per GNP to fit experimental measurements with a precision of $\pm 10$ molecules/GNP. $1030$ molecules/GNP fit well the experimental result, representing a surface density of only $0.229 \pm 0.002$ molecules$.nm^{-2}$. This density is very low compared to the expected $17$ molecules$.nm^{-2}$ deduced from the hydrodynamic radius or compared to Rostek et al. \cite{CitrateCompo} who find a surface density of $3.1$ molecules.$nm^{-2}$ with freshly prepared GNPs by Turkevich method. 
However this low estimated density is probably correct considering that our gold nanoparticles were washed several times contrary to the ones used by Rostek et .al \cite{CitrateCompo}.

As a consequence the previously described method should be very efficient and accurate to analyse ligand-shell density on nanoparticles using Geant4 and XPS analysis.

\section{Conclusion}

In this study we performed an XPS analysis of citrate-coated GNPs and gold bulk. We showed that XPS spectra ratios of GNPs to gold bulk is relevant to study GNPs electronic emission.
PENELOPE and Livermore-Geant 4 simulations of citrate-coated GNPs were undertaken for the first time, showing that the PENELOPE model does not well take into account photoelectric processes for low energy electrons whereas the Livermore model describes correctly the XPS experimental results. 

We demonstrated that the experimental electronic emission spectrum cannot be fully understood without considering the citrate-coating of GNPs and that Geant4 coupled to XPS measurements is a relevant tool to estimate the ligand density.

Finally, we observed an electronic emission enhancement above $300eV$ with GNPs when compared to gold bulk in the range of only a few percents. This "nano-scale effect" on electronic emission is not a coating effect but is intrinsic to GNPs and could participate to GNPs physical radio-sensitivity properties.

\bibliographystyle{plain}

\end{document}